\documentstyle[epsfig,preprint,aps,pra]{revtex}
\begin{document}
\tighten
\title{ Coherent Photoproduction of $\eta$-mesons on
        Three-Nucleon Systems }
\author{
        N. V. Shevchenko$^{1}$, V. B. Belyaev$^{1,2}$,
    S. A. Rakityansky$^{2}$, S. A. Sofianos$^2$, W.~Sandhas$^3$}
\address{$^1$Joint Institute  for Nuclear Research, Dubna, 141980,
         Russia}
\address{$^2$Physics Department, University of South Africa,
         P.O. Box 392, Pretoria 0003, South Africa}
\address{$^3$Physikalisches Institut, Universit\H{a}t Bonn,
         D-53115 Bonn, Germany}
\maketitle
%%%%%%%%%%%%%%%%%%%%%%%%%%%%%%%%%%%%%%%%%%%%%%%%%%%%%%%%%%%%%%%%%%%%%%%%
\begin{abstract}
A microscopic few-body description of near-threshold coherent
photoproduction of the $\eta$ meson on tritium and $^3$He
targets is given. The photoproduction cross-section is
calculated using the Finite Rank Approximation (FRA) of the
nuclear Hamiltonian.  The results indicate a strong final state
interaction of the $\eta$ meson with the residual nucleus.
Sensitivity of the results to the choice of the $\eta N$
$T$-matrix is investigated. The importance of obeying the
two-body unitarity condition in the $\eta N$ system is
demonstrated.

\end{abstract}
%%%%%%%%%%%%%%%%%%%%%%%%%%%%%%%%%%%%%%%%%%%%%%%%%%%%%%%%%%%%%%%%%%%%%%%%
\section{Introduction}
%%%%%%%%%%%%%%%%%%%%%%%

Investigations of the $\eta$-nucleus interaction are motivated
by various reasons. Some of them, such as the possibility of
forming quasi-bound states or resonances \cite{Heider} in the
$\eta$-nucleus system, are purely of nuclear nature. The others
are related to the study of the properties and structure of the
$S_{11}(1535)$ resonance which is strongly coupled to the $\eta
N$ channel.

For example, it is interesting to investigate the behavior of
the $\eta$-meson in nuclear media where, after colliding with
the nucleons, it readily forms the $S_{11}$ resonance.  The
interaction of this resonance with the surrounding nucleons can
be described in different ways \cite{Frank}, depending on
whether the structure of this resonance is defined in terms of
some quark configurations or by the coupling of meson-baryon
channels, as suggested in Ref. \cite{Keiser,Nieves}. The
estimation by Tiwari {\em et al.} \cite{Tiwari} shows, that
in case of pseudoscalar $\eta NN$ coupling there is an essential
density dependent reduction of the $\eta$-meson mass and of the
$\eta-\eta'$ mixing angle.

The importance of the influence of the nuclear medium on the
mesons passing through it, was recently emphasized by Drechsel
{\em et al.} \cite{Drech}. If this influence is described in
terms of self-energies and effective masses, then in the process
of $\pi$-meson passing through the nucleus, "saturation" of the
isobar propagator (or self-energy) takes place. This phenomenon
manifests itself even in light nuclei \cite{Drech}. Similar
ideas were discussed also in Ref. \cite{Fix1}. In other words,
the propagation of $\eta$-mesons inside the nucleus is a new
challenge for theorists.

Another interesting issue related to the $\eta$-nucleus
interaction, is the study of charge symmetry breaking, which may
partly be attributed to the $\eta-\pi^0$ mixing (see, for
example, Refs. \cite{Coon,Wilkin,Mag,Ceci}).  In principle, one
can extract the value of the mixing angle from experiments
involving $\eta$-nucleus interaction and compare the results
with the predictions of quark models.  However, to do such an
extraction, one has to make an extrapolation of the
$\eta$-nucleus scattering amplitude into the area of unphysical
energies below the $\eta$-nucleus threshold. This is a highly
model dependent procedure requiring reliable treatment of the
$\eta$-nucleus dynamics.

In this respect, few-body systems such as $\eta d$, $\eta$\,${}^3$He,
and $\eta$\,${}^4$He, have obvious advantages since they can be
treated using rigorous Faddeev-type equations. To the best of our
knowledge, so far only the simplest of these systems, namely the
$\eta(2N)$ system, was considered \cite{Ueda,Ours1,Fix2,Fix3} within
the exact AGS theory \cite{AGS}.

A solution of the few-body equations presupposes the knowledge of the
corresponding two-body $T$-matrices $t_{\eta N}$ and $t_{NN}$ off the
energy shell.  Due to the fact that at low energies the $\eta$ meson
interacts with a nucleon mainly via the formation of the
$S_{11}$-resonance, the inclusion of the higher partial waves
($\ell>0$) is unnecessary.  Furthermore, since the $\eta N$
interaction is poorly known, the effect of the fine tuned details of
the ``realistic'' $NN$ potentials would be far beyond the level of the
overall accuracy of the $\eta A$ theory.  Indeed, in contrast to the
well-established $NN$ forces, the $\eta N$ interaction is constructed
using very limited information available, namely, the $\eta N$
scattering length and the parameters of the $S_{11}$-resonance.
Furthermore, only the resonance parameters are known more or less
accurately while the scattering length (which is complex) is
determined with large uncertainties. Moreover, practically nothing is
known about the off-shell behavior of the $\eta N$ amplitude. It is
simply assumed that all mesons should have somewhat similar properties
and therefore the off-shell behavior of this amplitude could be
approximated (like in the case of $\pi$ mesons) by appropriate
Yamaguchi form-factors (see, for example,
Refs. \cite{Ueda,Ours1,Fix2,Fix3,Gar,Deloff}). However, if the
available data are used to construct a potential via, for example,
Fiedeldey's inverse scattering procedure \cite{Fied}, the resulting
form factor of the separable potential is not that simple.  The
problem becomes even more complicated due to the multichannel
character of the $\eta N$ interaction with the additional off-shell
uncertainties stemming from the $\pi$-meson channel.

In such a situation, it is desirable to narrow as much as possible the
uncertainty intervals for the parameters of $\eta N$ interaction.
This could be done by demanding consistency of theoretical
predictions based on these parameters, with  existing experimental
data for two-, three-, and four-body $\eta$-nucleus processes.  This
is one of the objectives of the present work. To do this, we calculate
the cross sections of coherent $\eta$-photoproduction on $^3$He and
$^3$H nuclei and study their sensitivity to the parameters of $\eta N$
amplitude.

%%%%%%%%%%%%%%%%%%%%%%%%%%%%%%%%%%%%%%%%%%%%%%%%%%%%%%%%%%%%%%%%%%%%%%
\section{Formalism}
%%%%%%%%%%%%%%%%%%%
We start by assuming that the Compton scattering on a nucleon,
$$
       \gamma + N \rightarrow N + \gamma\ ,
$$
as well as the processes of multiple re-appearing of the photon in the
intermediate states,
$$
   \gamma + N\rightarrow N+\eta\rightarrow \gamma +N\rightarrow
   N+\eta\rightarrow\dots\ ,
$$
give a negligible contribution to the
coherent $\eta$-photoproduction on a nucleus $A$. Then the process
\begin{equation}
\label{gAAeta}
        \gamma + A \rightarrow A + \eta\ ,
\end{equation}
can be formally described in two steps: at the first step, the
photon produces the $\eta$ meson on one of the nucleons,
\begin{equation}
\label{gNNeta}
        \gamma + N \rightarrow N + \eta\ ,
\end{equation}
and at the second step (final state interaction), the $\eta$ meson is
elastically scattered off the nucleus,
\begin{equation}
\label{etaAAeta}
        \eta + A \rightarrow A + \eta\ .
\end{equation}
An adequate treatment of the scattering step is, of course, the
most difficult and crucial part of the theory. Among the
few-body systems $\eta d$, $\eta$\,${}^3$H, $\eta$\,${}^3$He,
and $\eta$\,${}^4$He, so far only the simplest three-body one
($\eta d$) was considered in the framework of exact Faddeev-type
AGS equations.  The first  microscopic calculations concerning
the low-energy scattering of $\eta$-meson from ${}^3$H,
${}^3$He, and ${}^4$He nuclei were done in
Refs.~\cite{Ours2,Ours21,Ours22,Ours23,Ours24,Ours25,Ours26}
where the few-body dynamics of these systems were treated by
employing the Finite-Rank Approximation (FRA) \cite{FRA} of the
nuclear Hamiltonian. This approximation consists in neglecting
the continuous spectrum in the spectral expansion $$
   H_A=\sum_n{\cal E}_n|\psi_n\rangle\langle\psi_n|+
       \mbox{continuum}
$$
of the Hamiltonian $H_A$ describing the nucleus. Since the three- and
four-body nuclei have only one bound state, FRA reduces to
\begin{equation}
\label{FRAapprox}
   H_A\approx{\cal E}_0|\psi_0\rangle\langle\psi_0|\ .
\end{equation}
Physically, this means that we exclude the virtual excitations
of the nucleus during its interaction with the $\eta$ meson. It
is clear that the stronger the nucleus is bound, the smaller is
the contribution from such processes to the elastic $\eta A$
scattering.  By comparing with the results of the exact AGS
calculations, it was shown\cite{AGSetad} that even for $\eta d$
scattering, having the weakest nuclear binding, the FRA method
works reasonably well, which implies that we obtain sufficiently
accurate results by applying this method to the $\eta$\,${}^3$H,
$\eta$\,${}^3$He, and even more so to the $\eta$\,${}^4$He
scattering.

In essence, the FRA method can be described as follows (for details see
Ref.\cite{FRA}). Let
$$
    H=h_0+V+H_A
$$
be the total $\eta A$ Hamiltonian, where $h_0$ describes free
$\eta$-nucleus motion and
$$
     V=\sum_{i=1}^AV_i
$$
the sum of the two-body $\eta$-nucleon potentials. The
Lippmann-Schwinger equation
\begin{equation}
\label{LSinitial}
   T(z)=\sum_{i=1}^AV_i+\sum_{i=1}^AV_i(z-h_0-H_A)^{-1}T(z)
\end{equation}
for the $\eta$-nucleus $T$-matrix can be rewritten as
\begin{equation}
\label{Teq}
   T(z)=W(z)+W(z)M(z)T(z)\ ,
\end{equation}
where
\begin{eqnarray}
\label{kernel}
   M(z) &=& G_0(z)H_AG_A(z)\ ,\\
\label{G0}
   G_0(z) &=& (z-h_0)^{-1}\ ,\\
\label{Ga}
   G_A(z) &=& (z-h_0-H_A)^{-1}\ ,
\end{eqnarray}
and the auxiliary operator $W(z)$ is split into $A$ components of
Faddeev-type,
\begin{equation}
\label{T0sum}
   W(z) = \sum_{i=1}^AW_i(z)\ ,
\end{equation}
satisfying the following system of equations
\begin{equation}
\label{T0i}
   W_i(z) = t_i(z)+t_i(z)G_0(z)\sum_{j\ne i}^AW_j(z)\\
\end{equation}
with $t_i$ being the two-body $T$-matrix describing the interaction
of the
$\eta$-meson with the $i$-th nucleon, {\it i.e.}
\begin{equation}
\label{tetaNi}
   t_i(z) = V_i+V_iG_0(z)t_i(z)\ .
\end{equation}
It should be emphasized that up to this point no approximation
has been used yet and therefore the set of equations
(\ref{Teq}-\ref{tetaNi}) is equivalent to the initial
equation~(\ref{LSinitial}). However, to solve Eq. (\ref{Teq}),
we have to resort to the approximation (\ref{FRAapprox}) which
simplifies its kernel (\ref{kernel}) to
\begin{equation}
\label{kernelapprox}
    M(z)\approx
    \frac{{\cal E}_0|\psi_0\rangle\langle\psi_0|}
    {(z-h_0)(z-{\cal E}_0-h_0)}\ .
\end{equation}
With this approximation, the sandwiching of Eq. (\ref{Teq}) between
$\langle\psi_0|$ and $|\psi_0\rangle$ and the partial wave decomposition
give a one-dimensional integral equation for the
amplitude of the process (\ref{etaAAeta}).

The question then arises on how can a photon be included into
this formalism in order to describe the photoproduction process
(\ref{gAAeta}). This can be achieved by following the same
procedure as in Ref.\cite{Ours3} where the reaction
(\ref{gAAeta}) with $A=2$ was treated within the framework of
the exact AGS equations, and the photon was introduced by
considering the $\eta N$ and $\gamma N$ states as two different
channels of the same system. This implies that the operators
$t_i$ should be replaced by $2 \times 2$ matrices. It is clear,
that such replacements of the kernels of the integral
equations~(\ref{T0i}) and subsequently of the integral equation
(\ref{Teq}) lead to solutions having a similar matrix form
\begin{equation}
\label{matrices}
t_{i} \to \left(
\begin{array}{cc}
 t_i^{\gamma \gamma} & t_i^{\gamma \eta} \\
 t_i^{\eta \gamma}   & t_i^{\eta \eta}
\end{array}
\right) \Longrightarrow\quad
W_i \to \left(
\begin{array}{cc}
 W_{i}^{\gamma \gamma} & W_{i}^{\gamma \eta} \\
 W_{i}^{\eta \gamma}   & W_{i}^{\eta \eta}
\end{array}
\right)\Longrightarrow\quad
T \to \left(
\begin{array}{cc}
 T^{\gamma \gamma} & T^{\gamma \eta} \\
 T^{\eta \gamma}   & T^{\eta \eta}
\end{array}
\right)\ .
\end{equation}
Here $t_i^{\gamma \gamma}$ describes the Compton scattering,
$t_i^{\eta \gamma}$ the photoproduction process, and $t_i^{\eta \eta}$
the elastic $\eta$ scattering on the $i$-th nucleon.  What is finally
needed is the cross section
\begin{equation}
\label{sechenie}
      \frac{{\rm d}\sigma}{{\rm d}\Omega} =
      \frac{2}{9 (2 \pi)^2} \, \frac{k_{\gamma}}{k_{\eta}} \,
      \frac{E_\gamma m_A}{E_\gamma+m_A} \, \mu_{\eta A} \,
      \left|
      \langle\vec{k}_\eta,\psi_0|T^{\eta\gamma}({\cal E}_0+E_\gamma)
      |\psi_0,\vec{k}_\gamma\rangle
      \right|^2
\end{equation}
of the reaction (\ref{gAAeta}), where $\vec{k}_\gamma$ and $\vec{k}_\eta$
are the momenta of the photon and $\eta$ meson, $E_\gamma$ is the energy
of the photon, $m_A$ the mass of the nucleus, and $\mu_{\eta A}$ the reduced
mass of the meson and the nucleus.

However, it is technically more convenient to consider the
$\eta$-photoabsorption, {\it i.e.} the inverse reaction. Then
the photoproduction cross section can be obtained by applying
the detailed balance principle. The reason for this is that all
the processes in which the photon appears more than once, {\it
i.e.} the terms of the integral equations of type
$W^{\gamma\gamma}MT^{\gamma\eta}$ or
$W^{\eta\gamma}MT^{\gamma\eta}$ involving more than one
electromagnetic vertex, can be neglected.  Omission of these
terms in (\ref{Teq}) results in decoupling the elastic
scattering equation
\begin{equation}
\label{elastic}
      T^{\eta\eta}=W^{\eta\eta}+W^{\eta\eta}MT^{\eta\eta}
\end{equation}
from the equation for the photoabsorption
\begin{equation}
\label{absorption}
      T^{\gamma\eta}=W^{\gamma\eta}+W^{\gamma\eta}MT^{\eta\eta}\ .
\end{equation}
Once the $T^{\eta\eta}$ is calculated, the photoabsorption $T$-matrix
(\ref{absorption}) can be obtained by integration.

Therefore, the procedure of calculating the photoproduction
cross section (\ref{sechenie}) consists of the following steps:
\begin{itemize}
\item
Solving the system of equations
\begin{equation}
\label{proc1}
      W_i^{\eta\eta}=t_i^{\eta\eta}+t_i^{\eta\eta}G_0
      \sum_{j\ne i}^AW_j^{\eta\eta}
\end{equation}
for the auxiliary elastic-scattering operators $W_i^{\eta\eta}$\ .
\item
Calculating (by integration) the auxiliary matrices
$W_i^{\gamma\eta}$ from
\begin{equation}
\label{proc2}
      W_i^{\gamma\eta}=t_i^{\gamma\eta}+t_i^{\gamma\eta}G_0
      \sum_{j\ne i}^AW_j^{\eta\eta}\ .
\end{equation}
\item
Solving the integral equation
\begin{equation}
\label{proc3}
      T^{\eta\eta}=\sum_{i=1}^AW_i^{\eta\eta}+\sum_{i=1}^AW_i^{\eta\eta}
      MT^{\eta\eta}
\end{equation}
for the elastic-scattering $T$-matrix.
\item
Calculating (by integration) the photoabsorption $T$-matrix
\begin{equation}
\label{proc4}
      T^{\gamma\eta}=\sum_{i=1}^AW_i^{\gamma\eta}+
      \sum_{i=1}^AW_i^{\gamma\eta}MT^{\eta\eta}\ .
\end{equation}
\item
Substituting this $T$-matrix into Eq. (\ref{sechenie}) to obtain the
differential cross section for the photoproduction.  This is possible
because the absolute values of the photoproduction and photoabsorption
$T$-matrices coincide.
\end{itemize}

%%%%%%%%%%%%%%%%%%%%%%%%%%%%%%%%%%%%%%%%%%%%%%%%%%%%%%%%%%%%%%%%%%%%%%
\section{Two-body interactions}
%%%%%%%%%%%%%%%%%%%%%%%%%%%%%%%
To implement the calculation steps described in the previous section,
we need the two-body $T$-matrices $t^{\eta\eta}$ and $t^{\gamma\eta}$
for the elastic $\eta N$ scattering and the photoabsorption
$N(\eta,\gamma)N$ on a single nucleon respectively. Furthermore, all
equations (\ref{proc1}-\ref{proc4}) have to be sandwiched between
$\langle\psi_0|$ and $|\psi_0\rangle$ (ground state wave function of
the nucleus).  Since at low energies both the elastic scattering and
photoproduction of $\eta$ meson on a nucleon proceed mainly via formation
of the $S_{11}$ resonance, we may retain only the $S$-waves in the
partial wave expansions of the corresponding two-body $T$-matrices.

%%%%%%%%%%%%%%%%%%%%%%%%%%%%%%%%%%%%%%%%%%%%%%%%%%%%
\subsection{Elastic $\eta N$ scattering}
%%%%%%%%%%%%%%%%%%%%%%%%%%%%%%%%%%%%%%%%%%%%%%%%%%%%
The problem of constructing an $\eta N$ potential or directly the
corresponding $T$-matrix $t^{\eta\eta}$ has no unique solution since
the only experimental information available consists of the
$S_{11}$-resonance pole position $E_0-i\Gamma/2$ and the $\eta N$
scattering length $a_{\eta N}$.  In the present work, we use three
different versions of $t^{\eta\eta}$.

%%%%%%%%%%%%%%%%%%%%%%%%%%%%%%%%%%%%%%%%%%%%%%%%%%%%%
\subsubsection{Version I}
%%%%%%%%%%%%%%%%%%%%%%%%%%%%%%%%%%%%%%%%%%%%%%%%%%%%%
With the absence of any scattering data it is practically impossible
to construct a reliable $\eta N$ potential. However, we can make use
of the fact that the $S_{11}$-resonance is the dominant feature of the
$\eta N$ interaction in the low-energy region, where the elastic
scattering can be viewed as the process of formation and subsequent
decay of this resonance, {\it i.e.}
\begin{equation}
\label{diagram}
        \eta + N\,\longrightarrow\,S_{11}\,
        \longrightarrow\,N+\eta\ .
\end{equation}
This implies that in this region the corresponding Breit-Wigner formula
could be a good approximation for the $\eta N$ cross section.
Therefore, we may adopt the following ansatz
\begin{equation}
    t^{\eta\eta}(k',k;z) = g(k') \, \tau(z) \, g(k)
\label{tetan}
\end{equation}
where the propagator $\tau(z)$ describing the intermediate state of
the process (\ref{diagram}), is assumed to have a simple Breit-Wigner
form
\begin{equation}
    \tau(z) = \frac {\lambda}{z - E_0 + i\Gamma/2}\ ,
\label{tau1}
\end{equation}
which guaranties that the $T$-matrix (\ref{tetan}) has a pole at the
proper place.  The vertex function $g(k)$ for the processes $\eta
N$\,$\leftrightarrow S_{11}$ is chosen to be
\begin{equation}
\label{ffactor}
               g(k)=(k^2+\alpha^2)^{-1}
\end{equation}
which in configuration space is of Yukawa-type. The range parameter
$\alpha = 3.316$\,fm$^{-1}$ was determined in Ref.~\cite{alpha} while
the parameters of the $S_{11}$-resonance
$$
        E_0=1535\,{\rm MeV}-(m_N+m_\eta)\ ,
        \qquad \Gamma=150\,{\rm MeV}
$$
are taken from Ref.~\cite{PDG}.  The strength parameter
$\lambda$ is chosen to reproduce the $\eta$-nucleon scattering
length $a_{\eta N}$,
\begin{equation}
     \lambda= 2 \pi \, \frac{\alpha^4(E_0-i\Gamma/2)}
        {\mu_{\eta N}}a_{\eta N}\,.
\label{t000}
\end{equation}
the imaginary part of which accounts for the flux losses into the $\pi
N$ channel. Here $\mu_{\eta N}$ is the $\eta N$ reduced mass.

The two-body scattering length $a_{\eta N}$ is not accurately
known. Different analyses~\cite{Batinic} provided values for $a_{\eta N}$
in the range
\begin{equation}
           0.27\ {\rm fm}\le{\rm Re\,}a_{\eta N}\le 0.98\
{\rm fm}\ ,\qquad
       0.19\ {\rm fm}\le{\rm Im\,}a_{\eta N}\le 0.37\ {\rm fm}\ .
\label{interval}
\end{equation}
In most recent publications, the value used for Im $a_{\eta N}$
is around $0.3$\,fm.  However, for Re\,$a_{\eta N}$ the
estimates are still very different (compare, for example,
Refs.~\cite{Green97} and~\cite{Speth}).  In the present work we
assume that
\begin{equation}
\label{etaNlength}
        a_{\eta N} = (0.55 + i 0.30)\,{\rm fm}\ .
\end{equation}
The $T$-matrix $t^{\eta\eta}$ constructed in this way, reproduces the
scattering length (\ref{etaNlength}) and the $S_{11}$ pole, but apparently
violates the two-body unitarity since it does not obey the two-body
Lippmann-Schwinger equation.

%%%%%%%%%%%%%%%%%%%%%%%%%%%%%%%%%%%%%%%%%%%%%%%%%
\subsubsection{Version II}
%%%%%%%%%%%%%%%%%%%%%%%%%%%%%%%%%%%%%%%%%%%%%%%%%
An alternative way of constructing the two-body $T$-matrix
$t^{\eta\eta}$ is to solve the corresponding Lippmann-Schwinger
equation with an appropriate separable potential having the same
form-factors (\ref{ffactor}).  However, a one-term separable
$T$-matrix obtained in this way, does not have a pole at
$z=E_0-i\Gamma/2$.  To recover the resonance behavior in this case,
we use the trick suggested in Ref.~\cite{Deloff}, namely, we use an
energy-dependent strength of the potential
$$
       V(k,k';z)=g(k)\,
         \left[ \Lambda + C \frac{\zeta}{\zeta-z} \right] \, g(k')
$$
where $\Lambda$ is complex while $C$ and $\zeta$ are real constants.
With this ansatz for the potential, the Lippmann-Schwinger equation
gives the $T$-matrix in the form~(\ref{tetan}) with
\begin{equation}
    \tau(z) = -\left(\frac{4 \pi \alpha^3}{\mu_{\eta N}}\right)
    \frac{\Lambda (\zeta-z) + C \zeta}
    {\zeta-z-\left[\Lambda (\zeta-z)
    + C \zeta\right] / (1 - i\sqrt{2 z \mu_{\eta N}} /\alpha)^2}\ .
\label{tdeloff}
\end{equation}
The constants $\Lambda$, $C$, and $\zeta$ can be chosen in such
a way that the corresponding scattering amplitude reproduces the
scattering length $a_{\eta N}$ and has a pole at
$z=E_0-i\Gamma/2$.

This version of $t^{\eta\eta}$ also reproduces the scattering length
(\ref{etaNlength}) and the $S_{11}$ pole.  Moreover, it is consistent
with the condition of the two-body unitarity.

%%%%%%%%%%%%%%%%%%%%%%%%%%%%%%%%%%%%%%%%%%%%%%%%%
\subsubsection{Version III}
%%%%%%%%%%%%%%%%%%%%%%%%%%%%%%%%%%%%%%%%%%%%%%%%%
We can also construct the $t^{\eta\eta}$ having the same form as
version I, namely (\ref{tetan}), with the same $\tau(z)$ as in
(\ref{tau1}) but obeying the unitarity condition
\begin{equation}
\label{unitarity}
        (1-2\pi it^{\eta\eta})(1-2\pi it^{\eta\eta})^\dag=1\ .
\end{equation}
Of course, with the simple form (\ref{tetan}), we cannot satisfy
the condition (\ref{unitarity}) at all energies. To simplify the
derivations, we impose this condition on $t^{\eta\eta}$ at
$z=E_0$. Since Eq. (\ref{unitarity}) is real, it can fix only
one parameter and we need one more condition to fix both the
real and imaginary parts of the complex $\lambda$.  As the
second equation, we used the imaginary part of Eq.~(\ref{t000})
with $a_{\eta N}$ given by (\ref{etaNlength}).

This procedure guaranties the two-body unitarity and gives the correct
position of the resonance pole, but the resulting $t^{\eta\eta}$ gives
$a_{\eta N}$ which, of course, is different from the value
(\ref{etaNlength}), namely, it gives
\begin{equation}
\label{etaNlength1}
     a_{\eta N}=(0.76 +i0.61)\,{\rm fm}\ .
\end{equation}

In what follows we use the three versions of the matrix $t^{\eta\eta}$
described above. All of them have the same separable form
(\ref{tetan}) but different $\tau(z)$.  A comparison of the results
obtained with these three $T$-matrices can give us an indication of the
importance of the two-body unitarity in the photoproduction
processes.

%%%%%%%%%%%%%%%%%%%%%%%%%%%%%%%%%%%%%%%%%%%%%%%%%%%%
\subsection{Photoabsorption $N(\eta,\gamma)N$}
%%%%%%%%%%%%%%%%%%%%%%%%%%%%%%%%%%%%%%%%%%%%%%%%%%%%
In constructing the photoabsorption $T$-matrix $t^{\gamma\eta}$, the
$S_{11}$ dominance in the near-threshold region also plays an
important role. It was experimentally shown~\cite{Krusche} that, at
low energies, the reaction (\ref{gNNeta}) proceeds mainly via
formation of the $S_{11}$-resonance, {\it i.e.}
\begin{equation}
\label{diagramgamma}
        \gamma + N\,\longrightarrow\,S_{11}\,
        \longrightarrow\,N+\eta\ .
\end{equation}
This implies that $t^{\gamma
\eta}$ in this energy region can be written in a separable form
similarly to (\ref{tetan}). To construct such a separable $T$-matrix,
we use the results of Ref.~\cite{Green99} where the $t^{\gamma \eta}$
was considered as an element of a multi-channel $T$-matrix which
simultaneously describes experimental data for the processes
\begin{eqnarray}
\nonumber
    &{}& \pi + N \to \pi + N,    \quad   \pi + N \to \eta + N,
\\
\nonumber
    &{}& \gamma + N \to \pi + N, \quad  \gamma + N \to \eta +
N
\end{eqnarray}
on the energy shell in the $S_{11}$-channel. In the present work, we
take the $T$-matrix $t_{\rm on}^{\gamma \eta}(E)$  from Ref.~\cite{Green99}
and extend it off the energy shell via
\begin{equation}
    t_{\rm off}^{\gamma \eta}(k',k;E) =
    \frac{\kappa^2 + E^2}{\kappa^2 + {k'}^2} \,
        t_{\rm on}^{\gamma \eta}(E) \,
    \frac{\alpha^2 + 2 \mu_{\eta N} E}{\alpha^2 + k^2}\ ,
\label{toff}
\end{equation}
where $\kappa$ is a parameter. The Yamaguchi form-factors used in this
ansatz, go to unity on the energy shell.  Since $\kappa$ is not known,
this parameter is varied in our calculations within a reasonable
interval $1\,{\rm fm}^{-1}<\kappa<10\,{\rm fm}^{-1}$ which is a typical
range for meson-nucleon forces. It is known that $t^{\gamma \eta}$ is
different for neutron and proton.  In this work we assume that they
have the same functional form (\ref{toff}) and differ by a constant
factor,
$$
       t_{\rm n}^{\gamma \eta} = A \, t_{\rm p}^{\gamma\eta}\ .
$$
Multipole analysis~\cite{Muk} gives for this factor the
estimate $ A= -0.84 \pm 0.15$\,.

%%%%%%%%%%%%%%%%%%%%%%%%%%%%%%%%%%%%%%%%%%%%%%%%%%%%
\subsection{Nuclear subsystem}
%%%%%%%%%%%%%%%%%%%%%%%%%%%%%%%%%%%%%%%%%%%%%%%%%%%%
Since the $T$-matrices $t^{\eta\eta}$ and $t^{\gamma\eta}$ are poorly
known and their uncertainties significantly limit the overall accuracy
of the theory, it is not necessary to use any sophisticated
(``realistic'') potential to describe the $NN$ interaction. Therefore
we may safely assume that the nucleons interact with each other only
in the $S$-wave state.

To obtain the necessary nuclear wave function $\psi_0$, we solve the
few-body equations of the Integro-Differential Equation Approach
(IDEA)~\cite{idea1,idea2} with the Malfliet-Tjon potential~\cite{mt}.
This approach is based on the Hyperspherical Harmonic expansion method
applied to Faddeev-type equations. In fact, in the case of $S$-wave
potentials, the IDEA is fully equivalent to the exact Faddeev
equations.  Therefore, the bound states used in our calculations, are
derived, to all practical purposes, via an exact formalism.

%%%%%%%%%%%%%%%%%%%%%%%%%%%%%%%%%%%%%%%%%%%%%%%%%%%%%%%%%%%%%%%%%%%%%
\section{Results and discussion}
%%%%%%%%%%%%%%%%%%%%%%%%%%%%%%%%%%%%%%%%%%%%%%%%%%%%%%%%%%%%%%%%%%%%%
Figures 1-5 show the results of our calculations for the total cross section
$$
     \sigma = \int
              \left(
              \frac{{\rm d}\sigma}{{\rm d}\Omega}
              \right)
              \,{\rm d}\Omega
$$ of the coherent process (\ref{gAAeta}). The calculations were
done for two nuclear targets, $^3$H and $^3$He, using the three
versions of $t^{\eta\eta}$ described in the previous section.
The curves corresponding to these three $T$-matrices are denoted
by (I), (II), and (III), respectively.

As can be seen in Fig. 1, the two versions of $t^{\eta\eta}$,
(I) and (II), give significantly different results despite the
fact that both of them reproduce the same $a_{\eta N}$ and the
$S_{11}$-resonance. This indicates that the scattering of the
$\eta$ meson on the nucleons (final state interaction) is very
important in the description of the photoproduction process.
This conclusion is further substantiated when our curves are
compared to the corresponding points (triangles) calculated for
the $^3$He target in Ref.~\cite{Tiator} where the final state
interaction was treated using an optical potential of the first
order.  It is well-known that the first-order optical theory is
not adequate at the energies near resonances. This is the reason
why the calculations of Ref.~\cite{Tiator} underestimate
$\sigma$ near the threshold where with $a_{\eta
N}=(0.55+i0.30)\,{\rm fm}$ the systems $\eta$\,${}^3$H and
$\eta$\,${}^3$He have resonances \cite{Ours23}.

A significant differences between the corresponding curves (I)
and (II) in Fig. 1 imply that two-body unitarity is important as
well. To clarify this statement, we compare in Fig. 2 the
results corresponding to the three choices of $\tau(z)$ in
(\ref{tetan}). Surprisingly, the curves (II) and (III) almost
coincide despite the fact that they correspond to different
$a_{\eta N}$ while both obey the two-body unitarity condition.

The last three figures, Figs. 3, 4 and 5, show the dependence of the
results on the choices of the parameters $\kappa$ and $A = t_{\rm
n}^{\gamma \eta}/t_{\rm p}^{\gamma \eta}$. Since nothing is known
about $\kappa$, we assume $\kappa=\alpha$ as the basic value for
it. This can be motivated by the fact that both the elastic scattering
and photoabsorption (production) of the $\eta$ meson on the nucleon go
via formation of the same $S_{11}$ resonance.  This means that at
least one vertex, namely, $\eta N\leftrightarrow S_{11}$ should
be the same for both the elastic scattering and photoabsorption. To
find out how crucial the choice of $\kappa$ is, we did two additional
calculations with $\kappa=1\,{\rm fm}^{-1}$ and $\kappa=10\,{\rm
fm}^{-1}$ (see Fig. 3). We see that even with this wide variation the
curves are not far from each other especially in the immediate
vicinity of the threshold energy. Therefore the dependence on $\kappa$
is not very strong and the choice $\kappa=\alpha$ can give us a
reasonable estimate for the photoproduction cross section. Fig. 3 also
shows an interesting tendency of increasing $\sigma$ when the range of
the interaction becomes smaller (when $\kappa$ grows).

Figs. 4 and 5 for the $^3$H and $^3$He targets, respectively, show the
dependence of $\sigma$ on the choice of the parameter $A$. An
interesting observation here is that the cross section for $\eta$
photoproduction is more sensitive to this parameter with the tritium
rather than the $^3$He target. This means that between these two
nuclei, the tritium is a preferable candidate for a possible
experimental determination of the ratio $A$.

The cusp exhibited by all the curves at the threshold of total
nuclear break-up, reflects losses of the flux into the
non-coherent channel. In a sense, this is a reflection of the
four-body unitarity which the FRA equations are consistent with.
%%%%%%%%%%%%%%%%%%%%%%%%%%%%%%%%%%%%%%%%%%%%%%%%%%%%%%
\acknowledgements{ The authors gratefully acknowledge financial
support from the University of South Africa, the Division for
Scientific Affair of NATO (grant CRG LG 970110), and the
DFG-RFBR (grant 436 RUS 113/425/1).  One of the authors (V.B.B.)
wants to thank the Physikalisches Institut Universit\H{a}t Bonn
for its hospitality. }

%%%%%%%%%%%%%%%%%%%%%%%%%%%%%%%%%%%%%%%%%%%%%%%%%%%%%%

%%%%%%%%%%%%%%%%%%%%%%%%%%%%%%%%%%%%%%%%%%%%%%%%%%%%%%%%
%%%%%%%%%%%%%%%    FIGURES %%%%%%%%%%%%%%%%%%%%%%%%%%%%%
%%%%%%%%%%%%%%%%%%%%%%%%%%%%%%%%%%%%%%%%%%%%%%%%%%%%%%%%
\begin{figure}
\begin{center}
\unitlength=1mm
\begin{picture}(130,100)
\put(0,0){\epsfig{file=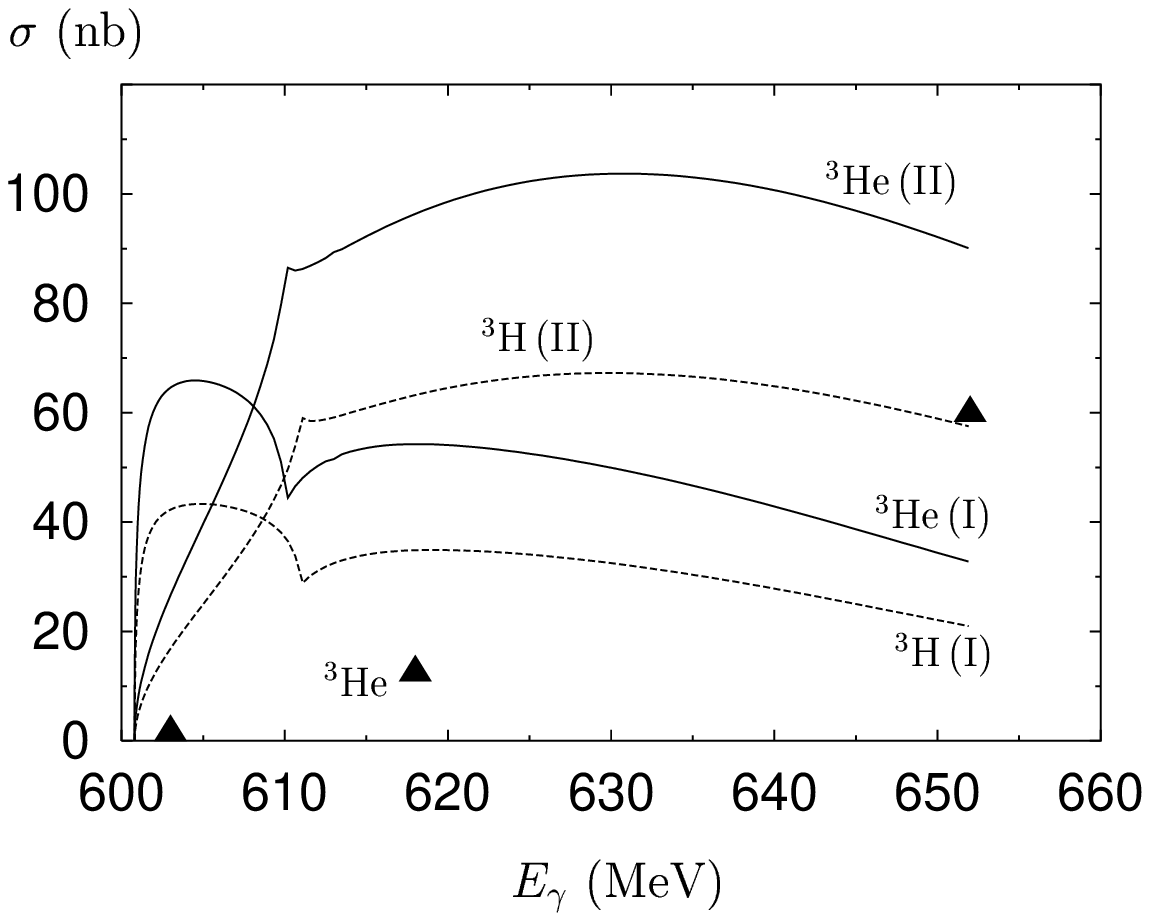}}
\end{picture}
\caption{
Cross section of the coherent $\eta$-photoproduction on the $^3$H and
$^3$He targets, calculated with the two versions of $t^{\eta\eta}$
which are denoted as (I) and (II) respectively. All curves correspond
to $a_{\eta N}=(0.55+i0.30)\,{\rm fm}$,
$\kappa=\alpha=3.316\,{\rm fm}^{-1}$, and $A=-0.84$\,.
The triangles represent the points calculated in Ref.
\protect\cite{Tiator} for
the $^3$He target within the optical model.
}
\label{fig1.fig}
\end{center}
\end{figure}
%%%%%%%%%%%%%%%%%%%%%%%%%%%%%%%%%%%%%%%%%%%%%%%%%%%%%%%%%%%
\newpage
\begin{figure}
\begin{center}
\unitlength=1mm
\begin{picture}(130,100)
\put(0,0){\epsfig{file=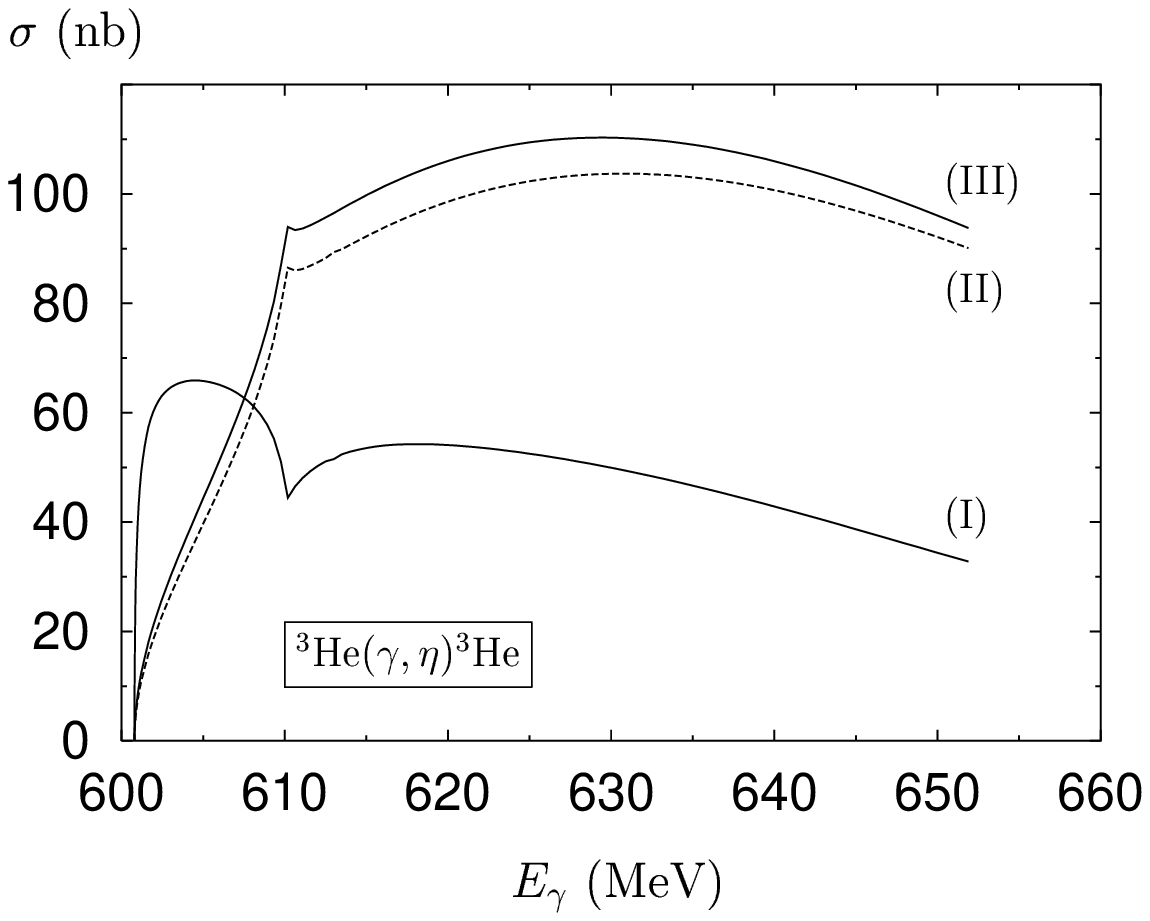}}
\end{picture}
\caption{
Cross section of the coherent $\eta$-photoproduction on $^3$He,
calculated with the three versions of $t^{\eta\eta}$
which are denoted as (I), (II), and (III) respectively.
All three curves correspond
to $\kappa=\alpha=3.316\,{\rm fm}^{-1}$  and $A=-0.84$\,.
For the curves (I) and (II) the $a_{\eta N}$ is given by
Eq.~(\ref{etaNlength}) while for the third curve by
Eq.~(\ref{etaNlength1})
}
\label{fig2.fig}
\end{center}
\end{figure}
%%%%%%%%%%%%%%%%%%%%%%%%%%%%%%%%%%%%%%%%%%%%%%%%%%%%%%%%%%%
\newpage
\begin{figure}
\begin{center}
\unitlength=1mm
\begin{picture}(130,100)
\put(0,0){\epsfig{file=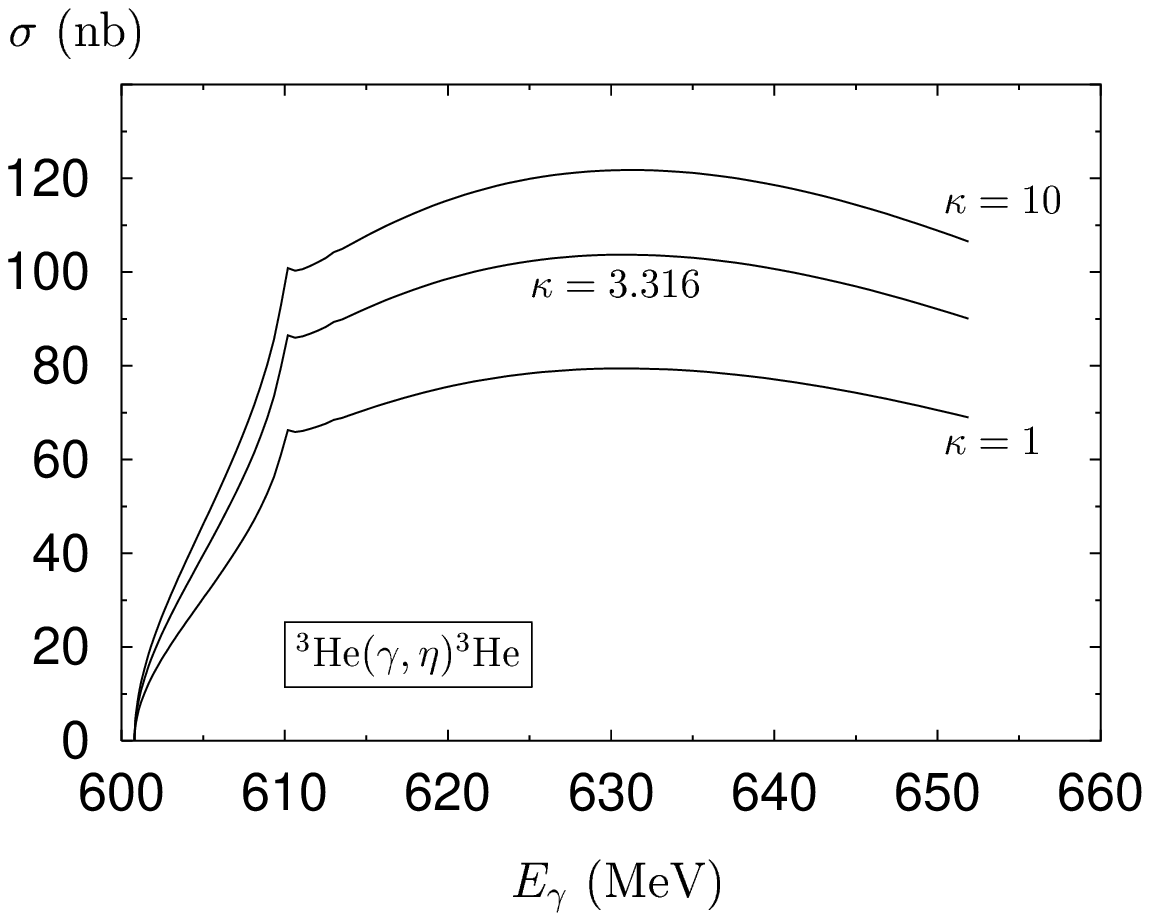}}
\end{picture}
\caption{
Cross section of the coherent $\eta$-photoproduction on $^3$He,
calculated with the version (II) of $t^{\eta\eta}$ with  three
values of the parameter $\kappa$. All three curves correspond
to $A=-0.84$\,.
}
\label{fig3.fig}
\end{center}
\end{figure}
%%%%%%%%%%%%%%%%%%%%%%%%%%%%%%%%%%%%%%%%%%%%%%%%%%%%%%%%%%%
\newpage
\begin{figure}
\begin{center}
\unitlength=1mm
\begin{picture}(130,100)
\put(0,0){\epsfig{file=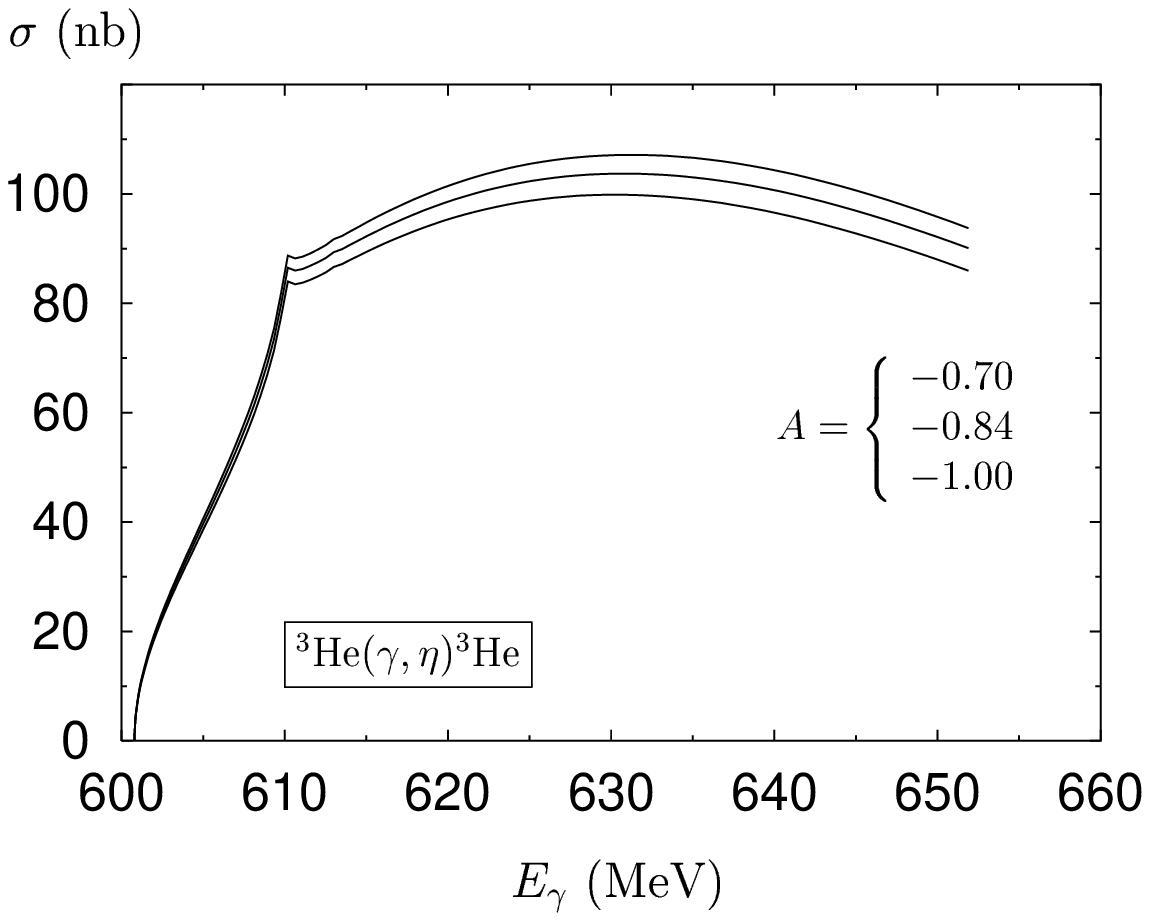}}
\end{picture}
\caption{
Cross section of the coherent $\eta$-photoproduction on $^3$He,
calculated with the version (II) of $t^{\eta\eta}$ with three different
values of the parameter $A$. All three curves correspond
to $\kappa=\alpha=3.316\,{\rm fm}^{-1}$\,.
}
\label{fig4.fig}
\end{center}
\end{figure}
%%%%%%%%%%%%%%%%%%%%%%%%%%%%%%%%%%%%%%%%%%%%%%%%%%%%%%%%%%%
\newpage
\begin{figure}
\begin{center}
\unitlength=1mm
\begin{picture}(130,100)
\put(0,0){\epsfig{file=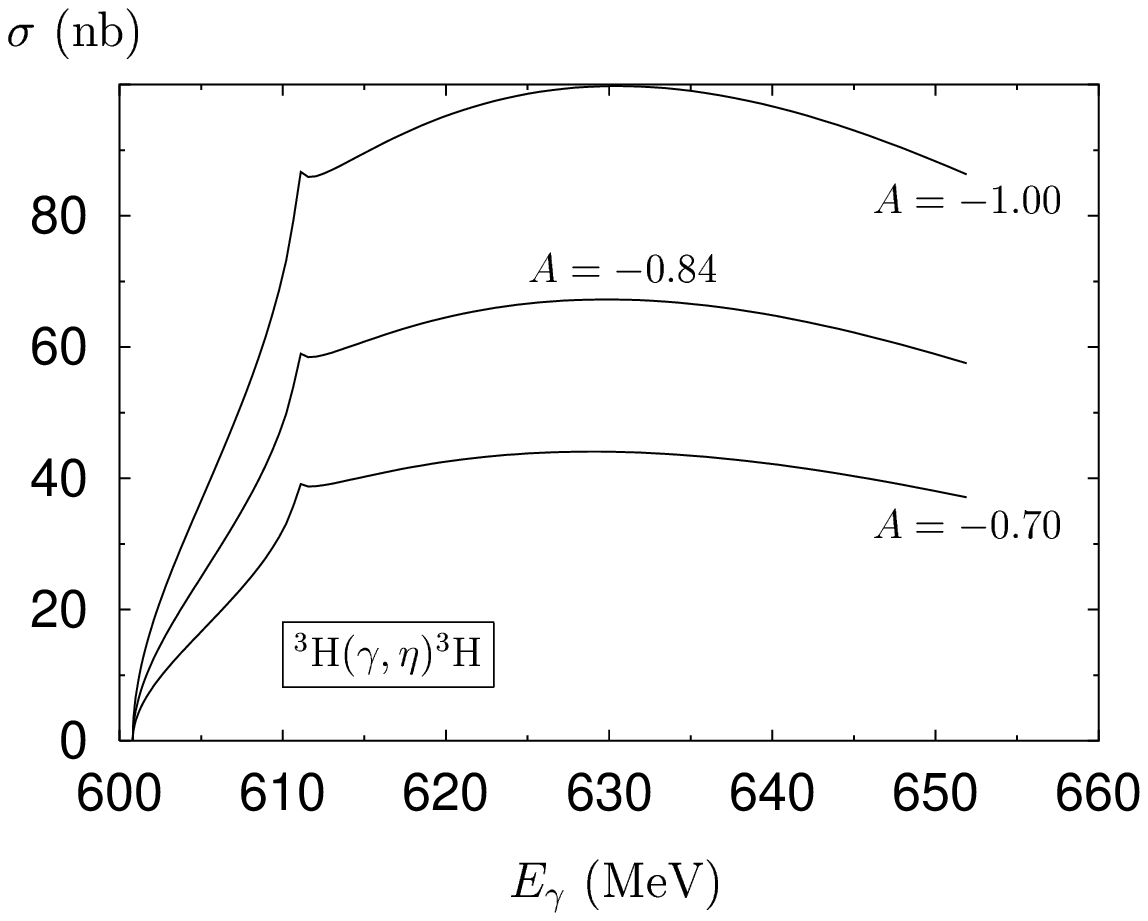}}
\end{picture}
\caption{
Cross section of the coherent $\eta$-photoproduction on $^3$H,
calculated with the version (II) of $t^{\eta\eta}$ with three different
values of the parameter $A$. All three curves correspond
to $\kappa=\alpha=3.316\,{\rm fm}^{-1}$\,.
}
\label{fig5.fig}
\end{center}
\end{figure}
%%%%%%%%%%%%%%%%%%%%%%%%%%%%%%%%%%%%%%%%%%%%%%%%%%%%%%%%%%%
\end{document}